# AR-based Modern Healthcare: A Review

Jinat Ara[1], Hanif Bhuiyan[2]*, Yeasin Aafat Bhuiyan[3], Salma Begum Bhyan[4],
Muhammad Ismail Bhuiyan[5]

**Abstract**

The recent advances of Augmented Reality (AR) in healthcare have shown that technology is a significant part of the current healthcare system. In recent days, augmented reality has proposed numerous smart applications in healthcare domain including wearable access, telemedicine, remote surgery, diagnosis of medical reports, emergency medicine, etc. The aim of the developed augmented healthcare application is to improve patient care, increase efficiency, and decrease costs. This article puts on an effort to review the advances in AR-based healthcare technologies and goes to peek into the strategies that are being taken to further this branch of technology. This article explores the important services of augmented-based healthcare solutions and throws light on recently invented ones as well as their respective platforms. It also addresses concurrent concerns and their relevant future challenges. In addition, this paper analyzes distinct AR security and privacy including security requirements and attack terminologies. Furthermore, this paper proposes a security model to minimize security risks. Augmented reality's advantages in healthcare, especially for operating surgery, emergency diagnosis, and medical training is being demonstrated here thorough proper analysis. To say the least, the article illustrates a complete overview of augmented reality technology in the modern healthcare sector by demonstrating its impacts, advancements, current vulnerabilities; future challenges, and concludes with recommendations to a new direction for further research.

# 1 Introduction

Augmented Reality (AR) stands for the comprehension of modern technology that performs through detection and identification of critical intuitive and provides competitive services. Nowadays Augmented reality is considered as one of the prominent advanced technology throughout the world [1]. The augmentation prototype enhanced its effectiveness and influenced almost every aspect of life such as emergency service, healthcare management, industrial implementation and many more [2, 3]. To generate more effective AR prototypes several researches have been undertaken and shown great achievements [4, 5]. According to Urakov, AR enhances the visualization of 3-dimensional holographic image through sophisticated glasses [6]. Again, El-hariri suggested that augmented reality is the visualization process of 3-D imagining data [7]. In some other studies, conducted by Vavra and Kim, augmented reality was described as a projection and interaction-based technology to interact and project computer-generated images in a real environment [8, 9]. Current AR related researches have resulted in several advanced prototypes, which promote AR as a prominent technology and call for future research to explore the unexplored prototypes and strategies [10, 11].

[1]Department of CSE, Jahangirnagar University, Savar, Dhaka, Bangladesh
[2] Data61 CSIRO, CARRS-Q, Queensland University of Technology, Queensland, Australia,
[3] Institute of Business Administration, Dhaka University, Dhaka, Bangladesh
[4] School of Science and Engineering, University of the Sunshine coast, Australia
[5] Sylhet MAG Osmani Medical College, Sylhet, Bangladesh
*Contact: hanif.bhuiyan@data61.csiro.au

These days, the advancement of AR technology in healthcare has become highly significant with its wide arrays of applications. AR-based healthcare systems are being incorporated with computer vision, object detection and identification, image processing, image segmentation [12] and cloud computing technology [13-16]. AR-based healthcare solution has achieved several prominent advancements, such as more secure connectivity and privacy across individual patients [17], more speed, while providing quality diagnosis & treatment with more reliability. Wireless services, early diagnosis, real-time monitoring, online-consultation, tumor detection, diagnosis specialization, m-health service and personal assessment are the emerging prototypes of AR-based healthcare solution as shown in Figure1. Similarly, Medication adherence and rehabilitation treatment have also enhanced through its utilization. Recently, modern healthcare society introduced e-health policy for the implementation of prominent augmented devices: AR headset and smart glasses has been recognized as remarkable advancements by healthcare professionals [18, 19].

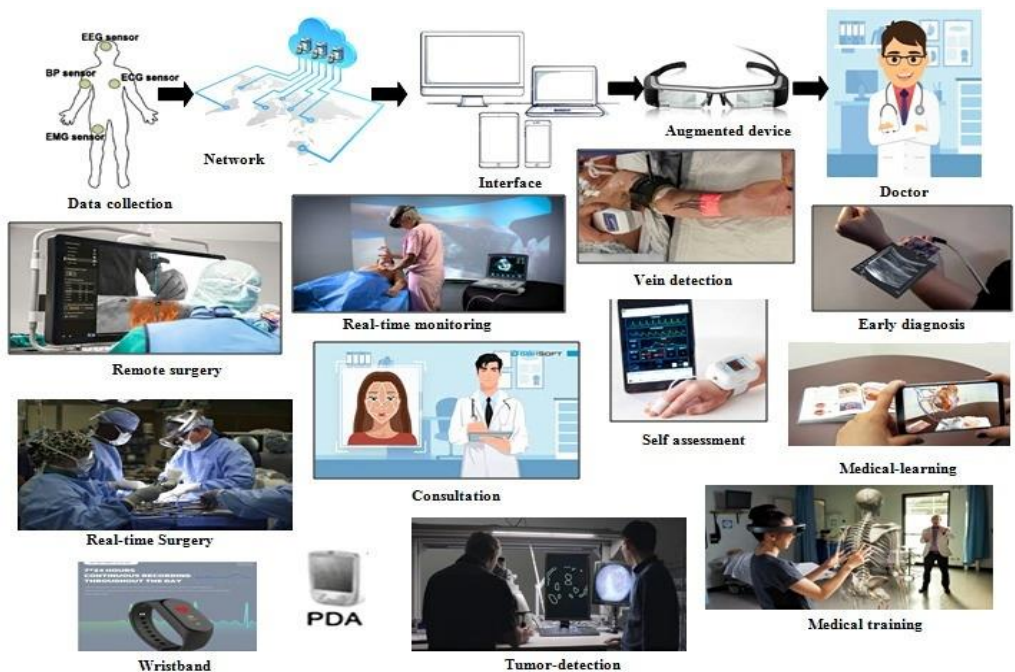

**Figure 1: Current trends of augmented reality in healthcare platform**

Because of its growing popularity, more and more works are being proposed towards its development. Majority of the proposed solutions developed are concerned with AR-based healthcare applications [20], some of which describe the potentiality and impacts of modern AR- healthcare applications. With keen observation, we concluded that the majority of proposed works study the development of AR-based m-health and clinical applications instead of their development strategy, advanced healthcare services, challenges and limitations [21, 22]. Although, a comprehensive explanation of the application security and functionalities could enhance the accuracy of future development approaches. By addressing these issues, this paper investigates the impacts of augmented reality in the healthcare system considering current healthcare trends, advanced services, current opportunities and limitations. This paper can be effective in having insights of current augmented-based healthcare applications. For a further study of AR and healthcare solutions, interested readers are being referred to [23-25].

This article has been organized as following: Several AR healthcare application services including mobile application and clinical approaches are demonstrated in section 2. Various aspects of the AR healthcare applications development structure being presented in section3.Several AR healthcare applications related current issues and security are being described in section 4 and 5 respectively. The article is being summed up with a through conclusion and a recommended direction of future research.

## 2 The AR Healthcare Services and Applications

Nowadays, AR technology has attracted significant attention to developing applications that are directly used by healthcare professionals and patients. AR-based healthcare solutions can be applied to a diverse array of fields. In this section, we are going to describe the AR service platform and then will describe the available services and applications.

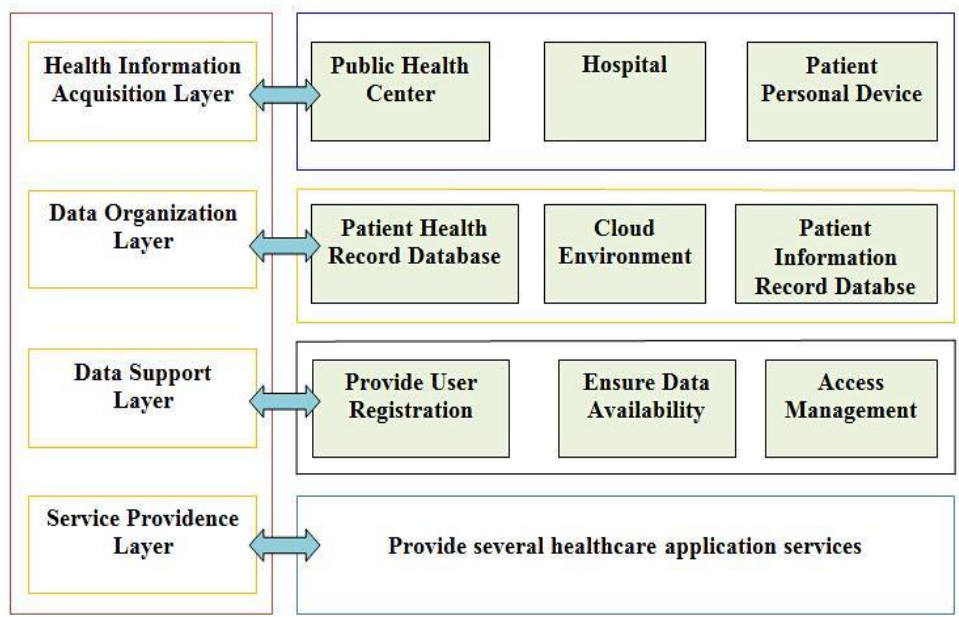

**Figure 2: The process of augmented healthcare application services**

The AR service platform refers to a framework that focuses on healthcare information. The health information shared by public health centers and some hospitals are being made available for the medical specialists or doctors. It provides services through: information acquisition layer, data organization layer, data support layer and service providence layer as shown in Figure 2. The health information acquisition layer refers to the organization of healthcare information or patient health report. Data organization layer focuses on various databases such as patient health record database or cloud environment. These databases provide services for healthcare applications during data examination or visualization. The data support layer provides services through user registration, data availability and accessing opportunity. To ensure authentication service and to reduce unknown threat it prerequisites the registration panel. Based on authenticity confirmation, it ensures data availability and accordingly grants access. Finally the service providence layer delivers the appropriate services. This way Augmented based healthcare application ensures the complete services. According to the context of healthcare, no standard definition of AR healthcare services can be found. As a result, it gets difficult to categorize these services. Still we have summarized the available AR healthcare applications and their services as following Table 1.

**Table 1: AR technology-based healthcare applications and services**

| Applications | Services |
|---|---|
| Eye diagnosis | -Identifies the patient's exact eye diseases: Glaucoma, dry eye - condition, etc.<br>-Using digital contact lens measures blood sugar levels through multi-sensor for retinal implanted people. |
| Cardiac treatment | -To understand cardiac inner and deeper structure.<br>-Analyzes the heart condition according to normal, murmur and extra-systolic sound.<br>-Cardiac data examination, prediction and cardiac arrhythmia treatment. |
| Cancer Detection | -Used to diagnosis breast cancer in sentinel lymph nodes and prostate cancer in prostatectomy specimens. |
| Brain Tumor Detection | -Identifying skin incision, skull craniotomy and tumor location. |
| Smart physical rehabilitation | -Provides inpatients and outpatient rehabilitation facilities for musculoskeletal, neurological, r-hematological and cardiovascular systems. |
| Surgical procedure | -Helps orthopedic surgeon to examine the abnormal joint function.<br>-Identify skin incision, skull craniotomy and tumor location for Choledochoscopy surgery.<br>-Prevents damaging tissues, blood vessels and dental nerves during dental surgery.<br>-Identifies the bone structure through 3D computed tomography (CT) data for orthopedic surgery.<br>-Helps in Endoscopic endonasal transsphenoidal surgery.<br>-Used for craniofacial surgery applied in treating orbital hypertelorism, hemifacial microsomia, mandibular angle split osteotomy related abnormalities. |
| Monitoring and guidelines | -Provides exact guidelines about healthy food.<br>-Provides information about allergic food, low-fat diets, and general caloric intake.<br>-Support for medication plan and medication restrictions.<br>-Provides feedback on lung conditions and used for the diagnosis of respiratory disease.<br>-Continuously monitors glucose level, insulin dosages and suggests appropriate foods.<br>-Helps hearing impaired people for museum visits. |

# 3   The AR Healthcare Application Development Strategies

Nowadays, stakeholders and software developers are interested to develop augmented healthcare solutions for its advanced prototypes and services. To develop advanced augmented healthcare application with full functionalities, development strategies are indeed needed. Focusing on this issue, in this section we are going to demonstrate the AR healthcare application development structure including effective communication and organization of numerous prototypes or entities. The development strategies are described through: data acquisition, data streaming, network or security, application development, and user interface as shown in Figure3.

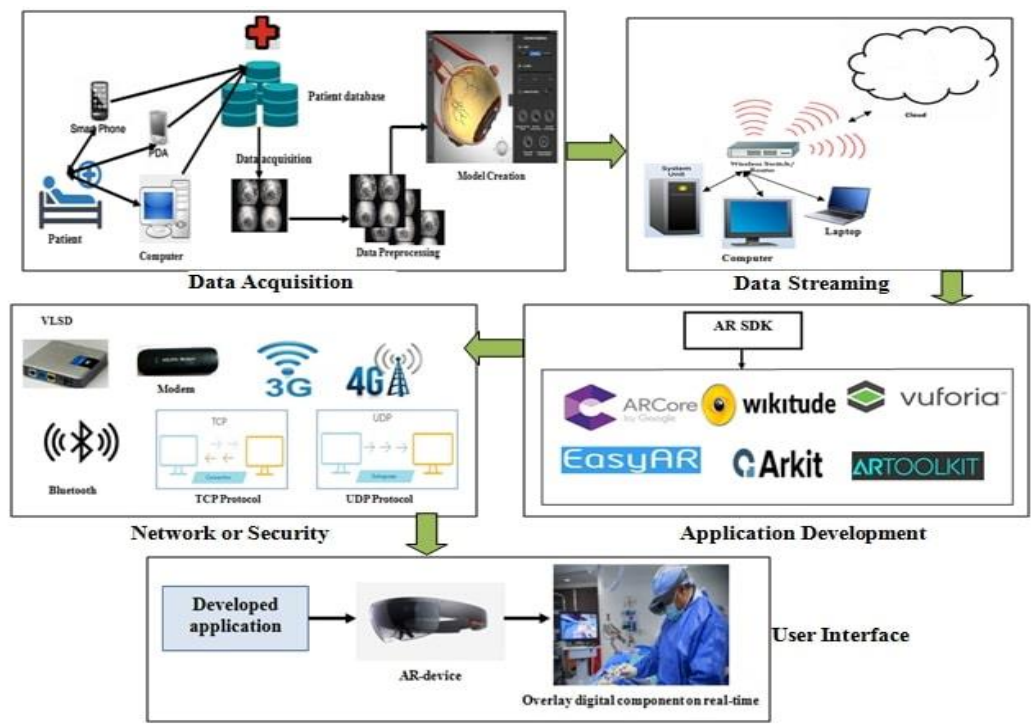

**Figure 3: Augmented healthcare application development interfaces**

## 3.1   Data Initializing

It is the primary layer of AR-based healthcare application development process. This layer supports data acquisition, preprocessing and model creation as shown in Figure 3. Data acquisition is the primary step in AR healthcare application development process. Generally, healthcare or hospital authority collects health data and stores them on the hospital database. Then data preprocessing is performed to reduce the noise in data and is performed on patient's data such as MRI, X-ray. Finally model creation helps to create a visual object for the effective organization of 3-D data [26]. This process allows the generation of a 3-D model from 2-D or 3-D health images or data. Some popular and user-friendly software for digital model creation are such as GLSL, ZBrush, AutoCAD, Blender, cinema4D, OpenGL and 3Ds studio max.

## 3.2 Data Streaming

Data streaming provides the benefits of transferring real-world information into a virtual environment such as cloud environment (Figure 5). Cloud environment or cloud server refers to the organization of information (health data) like files and the storage of these on the server [27].

## 3.3 Application Development

A framework, simply SDK, used for application development that provides a coding environment to define the functionality, several frameworks are being described in Table 2.

Table 2: Frequently used augmented healthcare application development frameworks

| Framework | Property and functionalities |
|---|---|
| ARKit | Apple developed the world's biggest augmented platform named ARKit. Motion tracking feature, back & front camera, generation of 2-D or 3-D content and integration on real-time make it more interactive and effective. Generally, ARKit is used for motion capturing, scene capturing, image processing and rendering. |
| ARCore | Google developed this SDK to provide services on iOS and android platforms. Motion tracking, environment understating and light estimating are the major functionalities of ARCore framework. 3-D holographic anatomy is the most significant healthcare application of the ARCore framework. |
| Wikitude AR | Wikitude AR framework is being developed for Android, iOS and Windows platforms. Objects tracking, overlaying images, scene capturing, geographical location tracking are some of the significant functionalities of this framework. |
| Vuforia | Vuforia supports Android, windows, iOS and smart glass. It assists to track images and 3D objects in real-time and overlay virtual 3D objects on real-world objects at the camera screen. Visualizing anatomical structure and detection of brain tumor through this make the surgical procedure easier and safer. |
| EasyAR | For android, iOS, Mac and Windows users, EasyAR is an effective framework. Image recognition, 3D object recognition, environment perception, cloud recognition, smart glass solution, screen recording, app cloud packaging, content and integration support are the prime functionalities of the EasyAR framework. |
| ARToolKit | ARToolKit is open-source augmented framework that supports iOS, Mac, android and windows users. Viewpoint tracking and virtual object interaction are the major functionalities of this SDK. |

## 3.4 Network Security

Network interface ensures security of numerous prototypes during communication. The entire prototype of AR healthcare application communicates through the network layer which ensures security through network platforms and network protocols (Figure 5). Network platform is the medium that generates frequencies for communication. Third-generation technology (3G) and fourth-generation technology (4G) are the most popular network platform or network technologies for AR-based healthcare applications. Managing data and ensuring secure connectivity are the prior responsibilities of network protocols. Two widely used protocols are User Datagram Protocol (UDP) and Transmission Control Protocol (TCP). UDP and TCP are both responsible for data transmission where UDP is connectionless; reliable and TCP, errorless.

## 3.5 User Interface

User interface allows the extraction of virtual environment data according to availability and accessibility. This layer also called the process of data collaboration. It helps healthcare specialist to overlay the virtually created imagining model on real-time patient's body using AR devices, shown in Figure 3.During surgery or diagnosis, on-demand health data accessing is the prime objective of this interface.

Majority of healthcare solutions are using smart display or glass. To detect object, image or text projection, Head Mounted Display and Holographic Display device are effective. HMD is being implemented in various healthcare solutions like, surgery, diagnosis, assessment, etc. Some prominent functionality of it includes recognizing voice comments, tracking eye movements and gestures movements. At present, the most powerful and prototype-based HMD devices are OHMDs and OST-HMD. Similarly, Holographic display is a lighting technology that visualizes a three-dimensional image of objects with better accuracy. Among several holographic displays, HoloLens and Oculus rift are the most used one in healthcare applications. In contrast, smart glass is a wearable hands-free control system that collects patient's physical information and provides accurate solutions during surgery and diagnosis. It helps to take pictures, record videos, recognize voice instruction and to lessen the requirement of looking on a different screen while examining medical test report [28]. Smart glasses are being categorized as optical glass and video glass for this study. It is worth mentioning that the majority of AR-based healthcare application uses optical glasses for its augmented functionality. The most prominent optical glasses are VuFine and Google Glass. Video glasses are called personal media viewers as they provide hands-free convenience. Video glasses are categorized into Vuzix, Epson Moverio and Atheer air.

# 4 The AR Healthcare Current Issues

One primal focus of this research is to propose solution for the technological and architectural problems associated with AR based healthcare. In addition, we noticed several other present and future possible challenges associated with augmented healthcare applications.

Figure 4 shows that, the most challenging issue of augmented platform is data security. To ensure data security, implementation of network connection encryption is essential. Ensuring completely encrypted or protected data transferring environment for healthcare applications is difficult but adequately needed. Another important issue is to launch a specialized platform that refers to applications package interface (APIs), framework, and appropriate libraries. It helps application designers and developers to manage effective code and classes, effective documentation, and creation of the template to build an AR-healthcare application. The Healthcare Leadership Council (HLC) published several guidelines related to healthcare IT interoperability. For accessing and sharing information, special security measures are a big concern. Generally, augmented healthcare application is the reflection of integrated digital components. These integrated digital components might introduce some technical barriers, especially for the healthcare specialist and the end-user. Overcoming the technical barrier is one of the major challenges. Furthermore, the development and implementation cost of the AR healthcare device is another important issue. Due to the high development cost, several healthcare providers disagree to reconstruct their healthcare platform. Also, a high range of implementation costs raises the barrier on introducing modern and digital technology into healthcare solutions. The current augmented platform is committed to provide a user-friendly environment with fluent functionality. But sometimes issues associated with improving consistency and reducing errors can be more viable. Overcoming this challenge may help to increase the device's acceptability. Developing an application with effective functionalities is a relatively focusable issue for AR technology.

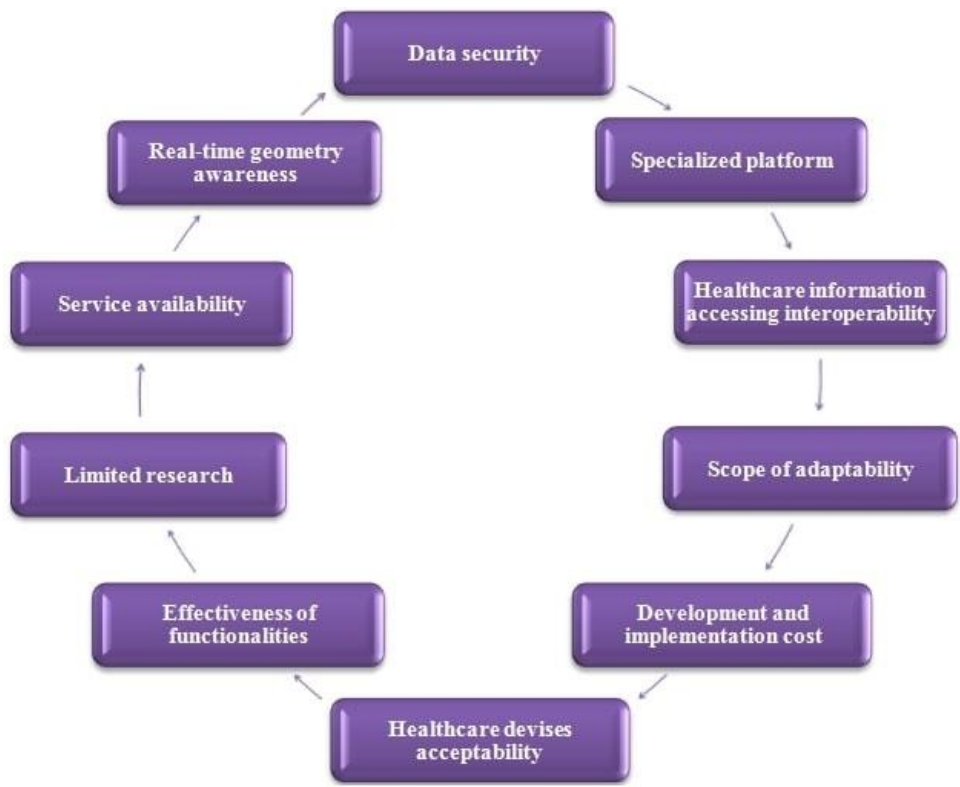

**Figure 4: Current issues of AR-based healthcare discipline**

Due to inappropriate prototypes, sometimes, unexpected abnormalities such as inappropriate recognition, lack of semantic meaning, and complex functionalities reduce the effectiveness of augmented healthcare applications. Consequently, for introducing effective prototypes and making the application more convenient, more effective future research his required. Instead of numerous services of AR application, providing constant services or long-duration services such as surgery visualization, projection, and interaction for as long as 9-12 hours in a constant manner is relatively difficult and challenging. Real-time geometry awareness is another major issue for researchers and developers. It associates with the process of annotation, tracking location, object detection, and appropriate measurement [29]. As majority of abnormalities arise during geometry transformation, awareness regarding real-time geometry can be a challenging issue for AR platform.

# 5 The AR Healthcare Security

Among several current AR healthcare issues, ensuring data security is one of the major concerns. Generally, augmented healthcare application provides services through a physical layer with overlaid computer-generated information [30]. Therefore, it is not free from external threat and even cyber security risk can be introduced on the physical layer. AR healthcare applications should address these external threats and common security vulnerabilities during the development phase. To facilitate the AR technology in healthcare domain, and to identify the security issues including security requirements, and vulnerabilities, a threat model is critical. This section discusses several security requirements, terminologies of external attacks, and provides a security model that might help prevent threats or network attacks.

## 5.1 Security Requirements

Security requirements of augmented healthcare applications refer to the property that protects the application platform from unexpected attack or threats. Figure 5 shows some security requirements that should be considered during application development to ensure application effectiveness and modality. For example, during the collaboration of virtual objects and real-time data, an intruder may attack. Augmented platform should provide security during the collaboration of virtual and real-time object. At present, the number of augmented healthcare devices has increased and is connected to the network. To provide medical data security, augmented platform should ensure confidentiality which forbids or prevent unregistered access to medical information. However, the assorted number of things can modify data during accessing or transmitting time. Healthcare solution should introduce integrity to ensure the security of medical information or health data that are not altered. Another important requirement is resiliency that ensures a protective environment if health devices are compromised. Another major security issue of current healthcare applications is availability. It ensures that data are stored and available in a secure environment and used by the registered authority under secure networks. To increase efficiency and to protect patient data from untraceable attack, ensuring the tracking functionality of all the activities having authorized detector is essential. For majority of the time, lack of awareness about DoS attacks can creates opportunities for malicious attackers to hamper data safety while transmitting. So security awareness should be focused. As we previously mentioned, augmented healthcare devices are wearable technologies, so they may connect to network anywhere. Hence, deriving security model for dynamic network topology is highly required. Designing a dynamic security protocol could increase the security of augmented applications. Surprisingly, augmented Reality Markup Language (ARML) has a lack of security control policy. Lack of security issues and Limited standardization make it vulnerable which in turn can introduce security vulnerabilities. In recent days, augmented reality is using computer-aided object provided by third-party vendors that paves way for threats such as spoofing, sniffing, data manipulation, and man-in-middle attacks.AR platforms should be aware of these threats and be sure of the authenticity of the generated content during transmission process to make the application reliable. As augmented healthcare application often uses various unprotected web browsers, it can lead to security breach too.

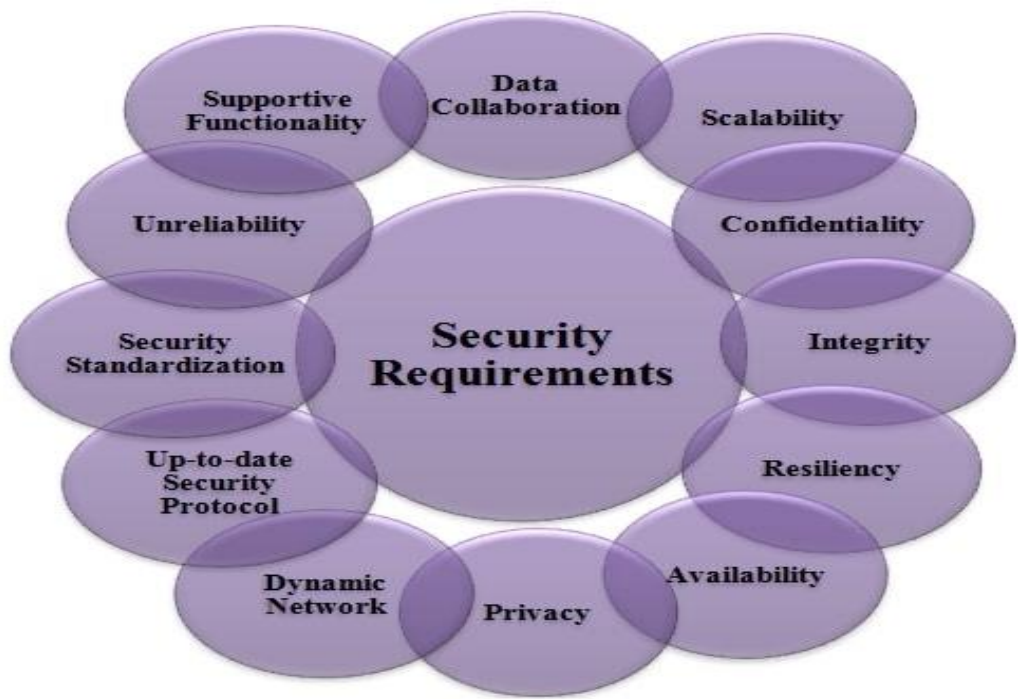

**Figure 5: Security requirements of AR platform**

## 5.2 External Attack Terminology

Augmented platform is not free from external attacks. Attackers may introduce several security threats and affect both existing and future augmented healthcare applications, devices, and networks. Sometimes these threats are predictable while some are hard to detect or predict. In this section, we are going to mention several external threats that reduce application and network effectiveness. Figure 6 shows some attack terminologies that can make systems vulnerable and reduce effectiveness. For example, by data-stealing or spreading malicious codes during data transmission or communication, intruders may replace original data or make it unreliable. Through unauthorized access attackers may modify the data and create confusion and reduces network performance. Also, false information may get injected to the network and reduce data reliability, affect hardware platform by altering program code, information or reprogramming. Sometimes attack can be introduced through software platforms such as operating systems or software applications and take advantage of software vulnerabilities that cause buffering, resource destruction or loss.

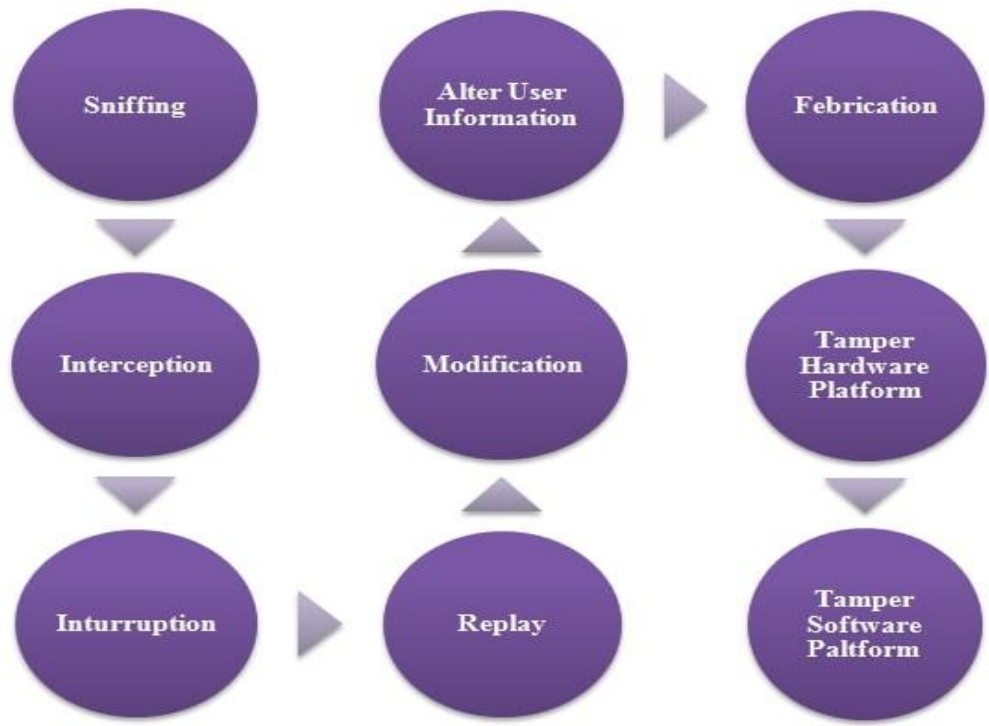

**Figure 6: Possible attacks for security threats**

## 5.3  The Proposed Security Model

Augmented reality-based developed healthcare applications and services are not yet robust but continuing to develop. Therefore, summing up the entire possible vulnerabilities, threats, and attacks associated with augmented medical solutions are difficult. With the expansion of healthcare devices, networks, and applications, a number of unknown or unseen threats may be initiated during communication or data transmission or data storage in the cloud.

To mitigate these unseen security issues especially in data transmission, security service should be designed with dynamic properties. An artificial intelligence-based dynamic algorithm might be capable of identifying these types of attacks. A security model for AR-based healthcare solutions has been proposed here. The proposed model operates through knowledge based services. Figure7 presents the proposed scheme that performs by following three security layers: threat detection layer, threat reduction layer and data protection layer. Threat detection layer is designed to receive healthcare data from healthcare devices and network and analyzes captured information to identify the presence of any threats. The threat reduction layer is designed to reduce the attack. Data protection layer helps protect data by defending all identified attacks. This security layer performs through dynamic algorithms and provides a shield against the attacks. Upon detection, threat detection layer issues action commands and shares detection experience with threat reduction layer. Threat reduction layer defends threat using threat redundant mechanism. Finally, data protection layer eliminates detected threats and protect the valuable data.

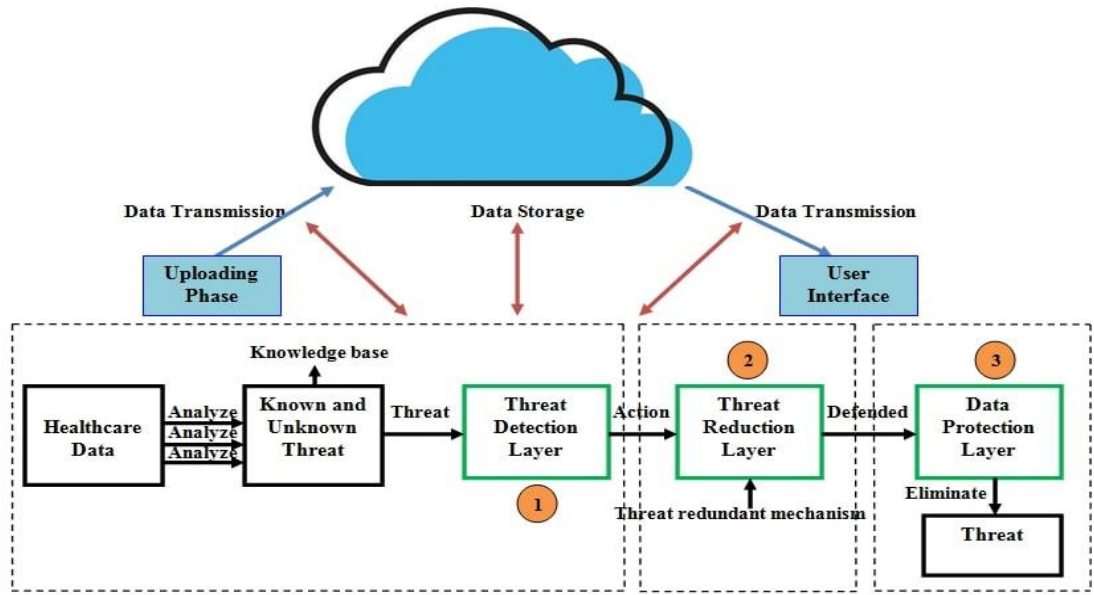

Figure 7: Proposed Intelligent Security model

# 6 Discussion

Throughout the world, researchers are vigorously working to invent better technological solution to enhance the modern health care system. The motive of these inventions is to bring a dramatic change in the healthcare sector and reduce the existing complexities. In this paper, a brief analysis was performed on various AR-based healthcare services and applications that deal with numerous medical data. A broad view of the recent development strategies regarding present healthcare application architecture and health data processing and accessing procedure has been laid out. The article, to some extent, tried to facilitate further development by pointing out several uncovered issues regarding concurrent security requirements and future challenges. The discussion performed in this article on standardization, data availability, service quality, and data protection may help in several ways for future research on AR-based healthcare applications and services. Moreover, this paper also illustrated the importance of augmented reality-based healthcare applications, backed up with present-day market data, which may increase the contribution of several stakeholders for further development. At the end, this review in many ways can be useful to researchers, health care professionals, policymaker, and several end-users working in the area of augmented technology and healthcare solution.